# SIMULATING THE COSMOS[1]: THE FRACTION OF MERGING GALAXIES AT HIGH REDSHIFT

P. Kampczyk[2], S. J. Lilly[2], C.M. Carollo[2], C. Scarlata[2], R. Feldmann[2], A. Koekemoer[3], A. Leauthaud[4], Y. Taniguchi[5], P. Capak[6]

[2] Institute of Astronomy, Department of Physics, ETH Zurich, CH-8093 Zurich, Switzerland
[3] Space Telescope Science Institute, Baltimore, Maryland
[4] Laboratoire d'Astrophysique de Marseille, France
[5] Tohoku University, Tokyo, Japan
[6] California Institute of Technology, Pasadena, USA

Abstract

Simulations of nearby ($0.015 < z < 0.025$) SDSS galaxies have been used to reproduce as accurately as possible the appearance that they would have on COSMOS ACS images if they had been observed at $z \sim 0.7$ and $z \sim 1.2$. These simulations include surface brightness dimming, the effects of band-pass shifting, the changing spatial resolution, and the increased noise. By adding the SDSS galaxies to random locations in the COSMOS images, we also simulate the effects of chance superpositions of high redshift galaxies with unrelated foreground or background objects. We have used these simulated images, together with those of real COSMOS galaxies at these same redshifts, to undertake a "blind" morphological classification of galaxies to identify those that appear to be undergoing mergers and thus to estimate the change in merger fraction with redshift. We find that real mergers are harder to recognize at high redshift, and also that the chance superposition of unrelated galaxies often produces the appearance of mergers where in reality none exists. In particular, we estimate that $1.5 - 2.0\%$ of objects randomly added to ACS images are misclassified as mergers due to projection with unrelated objects, and, as a result, that 40% of the apparent mergers in COSMOS at z=0.7 are likely to be spurious. Correcting for these two competing effects, we find that the fraction of galaxies undergoing mergers increases as $(1+z)^{3.8\pm1.2}$ to $z \sim 0.7$ and that this trend appears to continue to $z = 1.2$. Merger candidates at $z \sim 0.7$ are bluer than the parent population, especially when the statistical effects of the chance projections are accounted for. The automated non-parametric measures of morphology from Scarlata et al. (2006) show that the underlying galaxies of our merger candidates are more asymmetric than the population as a whole, and are often associated with irregular morphology. Nevertheless, the majority (~60%) of the merger candidates appear to be associated with spiral galaxies although in this case we cannot correct for the effects of chance projections.

*Subject headings:* galaxies:evolution – galaxies:high-redshift – galaxies:interacting – galaxies:structure

1. INTRODUCTION

All studies of the evolution of the galaxy population over cosmic time must take into careful account the observational consequences of the different distances and redshifts involved. These effects are in principle well understood and include differing physical scales (in kpc) of the instrumental resolution and/or pixelation, the shifting rest-wavelength of the detected light, the cosmological surface brightness dimming, and the different surface densities of unrelated foreground and background galaxies. Fortunately the availability of multi-color images of relatively nearby galaxies from the Sloan Digital Sky Survey (SDSS; York et al. 2000) enables us to simulate quite accurately the appearance that these nearby, present-day, galaxies would have had, if they were to be observed in exactly the same way as we actually observe the distant, younger, Universe. They are thus an invaluable tool in assessing evolutionary changes within the galaxy population.

In this paper we use SDSS $g$-band images of a volume-limited sample of 1813 low redshift galaxies to simulate the appearance of the present-epoch Universe if we had observed it in the COSMOS program (Scoville et al. 2006) at $z = 0.7$, and use SDSS $u$-band images of the same galaxies to simulate it at $z = 1.2$. Since the volume of the COSMOS cone is small at low redshifts, $z < 0.2$, this simulation is also of use to extend the baseline of evolutionary studies in COSMOS by providing a zero-redshift benchmark of the galaxy population. Our simulated images have been used as a present-epoch reference in the COSMOS-based studies of the evolution of the luminosity function of morphologically selected galaxies (Scarlata et al. 2006a), of the evolution of the massive galaxy population (Scarlata et al. 2006b), and of the evolution of the size function of disk galaxies (Sargent et al. 2006).

Here we use the simulated images in a direct comparison with a large sample of real COSMOS galaxies at $z_{phot} \sim 0.7$ and 1.2, to assess the redshift evolution of the fraction of visually-classified galaxy mergers from the present epoch up to $z \sim 1$. This visual classification complements the analysis of pair counts in COSMOS that has been undertaken with photometric redshifts (Kartaltepe et al. 2006) and future dynamical studies that will utilize the kinematic information on pairs from zCOSMOS redshift survey (Lilly et al. 2006). Any statistical analysis that is based on a visual classification is undoubtedly subjective. Our approach in this work is to try to ensure that this subjectivity is at least independent of redshift. This is achieved both through use of the simulated images to remove the redshift dependent observational biases discussed above, and through a rigorous "blind" methodology to eliminate any personal biases on the part of the classifier.

The merging of dark matter haloes is one of the most fundamental features of the current Cold Dark Matter paradigm for structure formation. A similar hierarchical growth will take place also for the baryonic component. However, in this case, the complexity of baryonic physics and observational biases may mask the underlying simplicity of the dark matter assembly. Also, it should be noted that most of the hierarchical assembly that has occurred in the recent past (i.e. back to $z \sim 2$) has occurred on mass scales that are larger than those of individual galaxies. Stars that are presently found in a given galaxy must have formed either in situ, or in another distinct dark matter halo that subsequently merged, or in a burst of star-formation associated with a merger event. Depending on how the baryons have been converted into stars within the dark matter haloes during the process of mass assembly, which will depend on currently uncertain cooling and heating processes, the relative fraction of stars in these three categories in present-day galaxies, in high redshift galaxies, and as functions of galaxy mass and environment, may be significantly different.

As well as changing the average stellar population properties and the stellar mass of galaxies, merging and other violent interactions may play a major role in redistributing existing stars within the galaxies leading to morphological transformations, e.g. the creation of spheroids. Major star-bursts associated with mergers may also have far-reaching effects on the subsequent evolution of galaxies, e.g. by consuming or expelling the bulk of the interstellar medium. Almost all of the most luminous galaxies in the present-day Universe appear to be associated with major mergers of two galaxies of comparable mass (e.g. Sanders & Mirabel 1996).

For all of these reasons, an observational determination of the rate at which galaxies are merging as a function of cosmic epoch and environment is a major, and at present rather uncertain, part of our view of galaxy formation and evolution. Four general approaches have been used to tackle this problem: (a) the study of close pairs of galaxies, preferably with kinematic information on their relative velocities, to assess the likelihood that they will merge in the near future; (b) the study of highly disturbed galaxies that appear to be actually merging; (c) the study of indirect indicators of on-going merging such as ultra-luminous star-burst galaxies or quasars, and (d) statistical reconstruction of the merging history via analysis of the galaxy mass function.

Most of these approaches are associated with measuring a fraction of galaxies merging or a total comoving number density of merging systems. To convert to an actual merger *rate* requires knowledge of the timescale of the merger event, or rather of the time that it would be recognized as such, which is rather uncertain. Thus, most observational studies of galaxy merging in the distant Universe have focused on the change in the merging fraction, parameterizing this as $(1+z)^m$, which will also be the change in the rate if the merging timescale is independent of redshift. There are many estimates of the exponent *m* in the literature and unfortunately, these are quite divergent in the derived value of *m*. While N-body cold dark matter simulations predict an increase of major merger rates of CDM haloes with redshift with $2.5 < m < 3.5$ (Governato et al. 1999; Gottlöber et al. 2001), different methods based on observations yield very discrepant estimates ranging from no or weak evolution in redshift ($m \sim 0$) to high evolution ($m \sim 6$), see e.g. Neuschaefer et al. 1995, Carlberg et al. 1994, Carlberg et al. 2000, Le Fèvre et al. 2000, Patton et al. 2002, Lin et al. 2004, Conselice et al. 2003, Lavery et al. 2004, Kartaltepe et al. 2006, Lotz et al. 2006. Some of these discrepancies may arise from using different criteria for a merger. For instance, kinematic studies usually require mergers of two roughly equal and relatively luminous galaxies, so that both components are included in spectroscopic surveys – see the discussion about the low evolution of kinematic close pairs in Berrier et al. 2006. However, some discrepancies persist between studies using apparently quite similar methodologies.

The large COSMOS survey offers a number of advantages for studying the merger history of galaxies. First, it covers a very large area (approximately 1.7 deg$^2$) with uniform depth ACS/HST images (Scoville et al. 2006, Koekemoer et al. 2006) and impressively deep multi-band ground-based images obtained with the Subaru and other telescopes (Taniguchi et al. 2006, Capak et al. 2006). Secondly, deep images are also being obtained in the X-ray (Hasinger et al. 2006), radio (Schinnerrer et al. 2006) and infrared Spitzer wavebands that produce high quality photometric redshifts and allow the identification of secondary indicators of mergers such as star-burst activity. Finally the COSMOS field is the subject of a major spectroscopic redshift survey (zCOSMOS, Lilly et al. 2006) that will produce accurate velocities $\sigma \ll 100$ kms$^{-1}$ for approximately 70% of the brighter galaxies ($I_{AB} < 22.5$) and characterization of the environments of galaxies out to high redshift. COSMOS will thus enable this problem to be approached from a number of complementary angles on the same region of sky, which should allow the clear disentangling of these observational issues. This paper addresses the question from the purely morphological point of view and will be supplemented in the future by others utilizing the broader COSMOS and zCOSMOS data sets. The reader is also referred to the pair analysis of Kartaltepe et al. (2006).

Throughout the paper we have assumed a concordance cosmology with $\Omega_\Lambda = 0.75$, $\Omega_M = 0.25$ and have adopted $H_0 = 70$ kms$^{-1}$Mpc$^{-1}$

2. SIMULATED GALAXIES

The aim of producing the simulated galaxies is to determine the appearance that local galaxies would have had if we had observed them at large distances and high redshifts as in the COSMOS HST/ACS images. It is relatively straightforward to simulate observational and cosmological effects like the shifting band-pass, the different pixel scales and point spread functions, and cosmological surface dimming, using the procedure described below. By pasting the artificially redshifted galaxies into real COSMOS ACS images, similar noise characteristics between the real and artificial images can also be obtained. Furthermore, by adding them at random locations (i.e. with no attempt to place simulated galaxies on regions of blank sky), the effects of superposition of unrelated foreground or background galaxies is also accurately simulated. The main subtlety is in the precise choice of the local sample to compare with the real high redshift galaxies, which is addressed in Section 3.2 below.

2.1   Selection of SDSS galaxies

The SDSS *g*-band well corresponds in central wavelength to the rest frame wavelength of objects observed in the ACS F814W filter at redshift $z \sim 0.7$. Therefore, band-width shifting effects ("*k*-corrections") can be eliminated if *g*-band SDSS images of local galaxies at $z \sim 0$ are used to produce simulated galaxies at $z \sim 0.7$. A similar correspondence applies between the SDSS *u*-band and F814W at a redshift $z \sim 1.2$. Of course, the width of these filters will not be precisely matched, but this will introduce only small second-order effects. These considerations fix the two redshifts to which the local SDSS galaxies will be artificially redshifted.

In order to fix a precise redshift range for the local SDSS sample, we need to consider the relative angular resolution of the COSMOS ACS and SDSS images, which we take to be 0.1 arcsec and 1.4 arcsec respectively (this may be slightly pessimistic for many SDSS galaxies but we have not tracked through the individual seeing measurements of galaxies), giving a angular resolution ratio of $S = 14$. In selecting local galaxies from SDSS, we therefore require that their angular diameter distances are *at least* 14 times smaller than the angular diameter distance to $z \sim 0.7$ (1509 Mpc), so that the spatial resolution in physical kpc is higher for the original SDSS galaxies than for the COSMOS galaxies at $z \sim 0.7$, and thus also at $z \sim 1.2$. This sets the maximum SDSS redshift to be $z = 0.028$.

Accordingly, a volume limited SDSS sample was constructed from the SDSS Data Release 4 (DR4) within the redshift range $0.015 \leq z \leq 0.025$ and $M_{B,AB} \leq -18.55$, motivating the absolute magnitude limits as described further below. About 25% of the SDSS galaxies were not readily usable for reasons that were unrelated to the galaxies' individual properties, e.g. because they were near to image boundaries. These were deleted from the original sample, leaving a statistically complete final sample of 1813 galaxies. We have checked that the distributions of absolute magnitude in the original and final sample are the same.

One of the goals of this exercise is to produce a zero-redshift benchmark for evolutionary studies using the very large COSMOS samples at high redshift, and in particular two distribution functions, the morphologically-selected luminosity functions (see e.g. Scarlata 2006b) and the size-functions (Sargent et al 2006). For these we need to know the number density of the SDSS sample. Direct computation of this would be quite tedious, requiring careful tracking of the spectroscopic sampling rate, redshift completeness, and the operational difficulties outlined above. This estimate would in any case be subject to cosmic variance etc. from the modest DR4 volume between 0.015 < z < 0.025, and it is therefore preferable to compute the effective volume based on the average density of the much larger full SDSS sample. Therefore, we have compared the (unnormalized) luminosity function of our 1813 galaxies with the absolute luminosity function of the full SDSS (Blanton et al. 2003) to derive an effective volume of our 1813-galaxy sample of $2.45 \times 10^5$ Mpc$^3$.

It is known that the SDSS spectroscopic sample has a bias against very bright objects with $r_{AB} < 14.5$. However the comparison of the distribution of luminosities in our sample with the luminosity function of the SDSS derived in Blanton et al. 2003 shows that the effect of incompleteness is very small in our sample.

2.2  Transforming SDSS galaxies into $z \sim 0.7$ and $z \sim 1.2$ COSMOS galaxies

For each SDSS galaxy at its actual redshift $z_i$, several simulated images were created at target redshifts $z_t = 0.7$ using the SDSS $g$-band images, and at $z_t = 1.2$ using the SDSS $u$-band images.

Because of the shallow number counts of stars in our own Galaxy, the stellar number density on the SDSS images is proportionally higher than on the COSMOS images. Thus, in order to avoid an apparent overdensity of stars around the SDSS galaxies in the simulated high redshift images, that might otherwise flag an SDSS galaxy and thereby defeat our blind methodology, we first removed the stars from the SDSS images. This was done with a semi-automated procedure using SExtractor (Bertin & Arnouts 1996). All images were inspected afterwards to check that this automatic removal of stars had not subtracted light from the galaxies themselves.

To transform the galaxies to high redshift, it is assumed that the PSF of both the ACS and SDSS images can be represented for the present purposes as a Gaussian function. We also assume that the seeing for SDSS images is a constant FWHM = 1.4 arcsec while the FWHM of the ACS images is taken to be 0.1 arcsec.

We therefore first smooth the relevant SDSS image of each galaxy with a flux-conserving Gaussian kernel, whose width is chosen so that, for each galaxy, the final (Gaussian) spatial resolution in kpc matches that of the ACS images at the target redshift. To be precise, we take the ratio of the gaussian widths σ of the point spread functions (PSFs) in the SDSS and ACS images to be the same for all objects, i.e.:

(1) $$S = \frac{\sigma_{SDSS}}{\sigma_{ACS}} \approx 14$$

We therefore require that the ratio of angular diameter distances $D_\theta$ at the input (as in SDSS) redshift, $z_i$ and at the target redshift to which they will be shifted, $z_t$ to be greater or equal to ratio of PSFs;

$$(2) \quad D_{i,t} = \frac{D_\theta(z_t)}{D_\theta(z_i)} \geq S$$

This then guarantees that the simulated SDSS galaxies at the target redshifts of 0.7 and 1.2 will have no worse physical resolution than the ACS images of real high redshift galaxies, allowing us to match the physical resolutions by degrading the SDSS images of the simulated objects.

$$(3) \quad X_{i,t} = \frac{D_{i,t}}{S} \geq 1$$

This increase in width of the point spread function is achieved by smoothing each galaxy image with a two-dimensional gaussian smoothing kernel (conserving total flux) that has a guassian σ that is given, for each galaxy, by $X_{i,t}$ (equation 3):

$$(4) \quad \sigma_{i,t} = \sigma_{SDSS}\sqrt{(X_{i,t}^2 - 1)}$$

The smoothed images are then re-pixelated (shrunk) through rebinning so as to match the COSMOS ACS pixel scale (0.05 arcsec), again conserving flux. Defining the ratio of pixel scales (0.4 and 0.05 arcsec for SDSS and COSMOS respectively) to be:

$$(5) \quad P = \frac{p_{SDSS}}{p_{COSMOS}} = 8$$

The rebinning factor is computed as:

$$(6) \quad R_{i,t} = \frac{D_{i,t}}{P} \geq 1.75$$

The rebinned images are then multiplied by a factor $F_{it}$ that accounts for the AB magnitude zero-point differences between the SDSS $g$ images and the COSMOS $I_{814}$ images (and between SDSS $u$ images and the COSMOS $I_{814}$ images) $ZP_{SDSS}$ - $ZP_{COSMOS}$, and for the $(1+z)^3$ surface brightness dimming term (appropriate for frequency flux density, $f_\nu$ since we are working in AB magnitudes) between $z_i$ and the target redshift $z_t$, i.e.

$$(7) \quad F_{i,t} = 10^{-0.4(ZP_{SDSS} - ZP_{COSMOS})} \frac{(1+z_i)^3}{(1+z_t)^3}$$

As discussed below (section 3.2), one question is whether to brighten the simulated galaxies to account for luminosity evolution between the target redshift and the present. Our pipeline produces images that are both unbrightened and images which are uniformly brightened by a factor corresponding to 0.7 magnitudes at $z = 0.7$ and by 1.2 magnitudes at $z$

= 1.2, i.e. Δμ = z. We refer to these below as the "unbrightened" and "brightened" images, respectively.

Finally, we checked that the noise per pixel in the processed SDSS images is well below the noise per pixel of the actual COSMOS ACS images, which it always is, by a factor of about 10. Although the SDSS images have been smoothed, the smoothing is sufficiently modest that the correlated noise from pixel to pixel is still much smaller than the pixel to pixel noise in the COSMOS images. The zodiacal background is approximately $\mu_{AB}$ = 22.0 mag arcsec$^{-2}$ and so we can assume that the COSMOS ACS images are background limited for these galaxies (20 < $I_{AB}$ <24), especially in the low S/N outer regions that are of most interest for many applications, including the present one. Therefore, we can simply add the manipulated images of SDSS galaxies to actual COSMOS ACS frames to obtain noise characteristics in the images of the added galaxies that are well matched to those of real faint COSMOS galaxies.

In order to simulate the effect of chance superpositions of foreground or background objects that may contaminate the images of high redshift galaxies, the added SDSS galaxies are positioned *randomly* in the ACS COSMOS tiles with respect to existing galaxies or image defects in the COSMOS ACS images, with no attempt made to locate them on regions of "blank" sky. For ease of cataloguing, the added images are added in batches of 49 galaxies in a precise seven by seven rectilinear grid separated by 10 arcsec. Although the input sample is limited to 1813 galaxies, each galaxy may be added to many different ACS images in order to increase the statistics of foreground and background contaminations. Each time, reflections or rotations of the added image may be applied to disguise the galaxy's identity.

At least to the eye, it is impossible to distinguish between the added SDSS objects and the real high redshift galaxies on the ACS images. The ACS tiles with added SDSS objects are available at IRSA along with explanatory documentation. An example is shown in Fig.1.

## 3. VISUAL IDENTIFICATION OF GALAXY MERGERS

The philosophy of our study has been to visually classify the images of both the real high redshift COSMOS galaxies and the added (simulated) SDSS galaxies, *without knowing whether each galaxy had come from COSMOS or SDSS*. This was achieved by extracting small "postage-stamps" of both real COSMOS galaxies and the simulated SDSS galaxies, shuffling them into a random (and unknown) order, and then visually and interactively inspecting them one by one. Specifically, the postage stamp images of: (a) real COSMOS galaxies at *z* ~ 0.7; (b) real COSMOS galaxies at *z* ~ 1.2; (c) added SDSS g-band galaxies at *z* = 0.7, both brightened and unbrightened; and (d) added SDSS u-band galaxies at *z* = 1.2 (brightened) were mixed together in a random sequence for the visual inspection. Both in design and in practice, it was impossible to tell, while classifying an individual object, to which of the above categories the object belonged to. Only in this way could we ensure that there were no conscious or subconscious redshift-dependent biases on the part of the classifier (PK). Even a "drift" over time in any biases present in the classification should not produce a redshift dependent effect/

### 3.1    Input samples from COSMOS at high z

To make a sensible comparison of the morphologies of the real COSMOS high redshift galaxies and of the added SDSS galaxies, the distribution of apparent $I_{AB}$ magnitude on the ACS images should be the same in both sets of objects.

The "real" high redshift galaxies are selected from a catalogue produced from the 260 COSMOS F814W ACS frames from the HST Cycle 12 observing period (Scoville et al. 2006). This catalogue was produced and used by Scarlata et al. (2006a), on the basis of a SExtractor-based compilation generated by Leathaud et al. (2006; see Scarlata et al. 2006a for details). The total area covered by the catalogue is 0.74 deg$^2$ and it contains approximately 56000 galaxies down to $I_{AB}$ = 24.0. Photometric redshifts for almost all of these galaxies have been computed using the COSMOS photometric data (Capak et al. 2006) and the ZEBRA photometric redshift code (Feldmann et al. 2006). For our study we have used the maximum likelihood photometric redshift from ZEBRA. Compared with zCOSMOS spectroscopic redshifts (Lilly et al. 2006) at $I_{AB}$ < 22.5, these show a small redshift dispersion at these magnitudes and at the redshifts of interest $z \leq 1.2$ ($\sigma_z/(1+z) \sim 0.03$) and a small (< 1%) catastrophic failure rate. We therefore use these photometric $z$ to isolate a set of galaxies with $0.6 < z_{phot} < 0.8$ and $1.18 < z_{phot} < 1.25$ to compare with the SDSS $g$-band and $u$-band images respectively. Although there are inevitably random errors in the photometric redshifts, the iterative ZEBRA code has a negligible systematic error, as tested with spectroscopic redshifts, and thus the use of photometric redshifts should not introduce a significant bias in our morphological study.

For the $0.6 < z < 0.8$ sample, the selection wavelength (i.e. the $I$-band at $z \sim 0.7$ and $g$-band at $z \sim 0$) is close enough to the rest-frame B-band that it is attractive to apply a simple selection in absolute B magnitude, $M_B \leq -19.25$. At the central redshift of this bin, this corresponds to an approximate limit of $I_{AB}$ < 23.25, i.e. 0.75 magnitudes above the limit of the parent catalogue. This ensures that a well-defined volume limited sample can be constructed across the redshift bin which spans all galaxy types. It should be noted that the absolute magnitude is computed from the ACS images (SExtractor magauto) and should not merge close pairs of galaxies into single objects, an effect which could artificially enhance the fraction of merging galaxies in any magnitude limited sample.

For the higher redshift $1.18 < z < 1.25$ sample we apply an $M_{B,AB} < -20.8$ selection. However, this is no longer complete in the sense that the bluest objects at $z \sim 0.7$ would have $I_{AB}$ > 24 (and are therefore not in the ACS sample). We address this below.

3.2 Input sub-samples from SDSS at low z

In order to generate comparison samples from SDSS with essentially the same distribution of apparent magnitudes as the high $z$ COSMOS samples, two approaches have been taken. The first is straightforward and is to apply exactly the same selection in absolute magnitude as at high redshift to the "unbrightened" SDSS images.

However, there is a great deal of evidence that the surface brightnesses of most types of galaxies (both disks and spheroids) are brighter at high redshift than at the present epoch by an amount that is approximately $\Delta\mu_B(mag) \sim \Delta z$ (Schade et al. 1995, Lilly et al. 1998, Barden et al. 2005). This is largely due to the passive evolution of stellar populations. Thus we have also used the "brightened" images of SDSS galaxies (see Section 2.2 above) coupled with a local selection that is fainter by the corresponding amount. The net effect is to produce an apparent $I_{AB}$ distribution of simulated SDSS galaxies at $z \sim 0.7$ that is very similar to that obtained using the unbrightened images (and similar to that in the real COSMOS data).

However, it does mean that the SDSS galaxies are selected from a different part of the luminosity function and this may have some consequences for the analysis if, e.g., the apparent merging fraction depends on luminosity, either directly (if the merging fraction at the present-epoch is a strong function of luminosity), or indirectly (if, e.g., the apparent sizes of galaxies depend on luminosity). In fact, we find such effects to be small.

Since, empirically, the appearance of galaxies does not depend strongly on luminosity, the overall appearance of the simulated galaxies does not depend much on whether we use the unbrightened (but intrinsically brighter) set of SDSS galaxies or the brightened objects (which extend to intrinsically fainter levels). However, it is not clear whether the morphological features that we actually use to identify mergers (e.g. tidal tails or other distortions) will have undergone this surface brightness evolution or not. If they are composed of stars of similar age to the bulk of the galaxy, then the surface brightness evolution is presumably appropriate. But if they are composed of young stars whose formation is triggered by the merger or interaction that is being observed, then they would have similar intrinsic surface brightness at all redshifts. This is our motivation for using both the SDSS galaxies with $M_{B,AB} < -18.55$ (brightened in the image processing by 0.7 mag) and those with $M_{B,AB} < -19.25$ (unbrightened) in our study. In practice, we find little change in our results whether we use the brightened or unbrightened samples; we are thus fairly confident that these subtleties are not influencing our results.

For the $z \sim 1.2$ simulations we only used $M_{B,AB} < -19.55$ (brightened) sample. To mimic the incompleteness with spectral type in the $z = 1.2$ COSMOS sample, mentioned above, we apply an additional criterion to the apparent magnitudes of the simulated galaxies at $z = 1.2$, namely $I_{AB} < 24.0$.

It should be noted that all SDSS absolute magnitudes are quoted as the Petrosian magnitudes in the SDSS DR4, measured *before* the objects are added to the ACS images. Therefore, like the real COSMOS photometry, their absolute magnitudes will not reflect the additional light from any superposed objects that were already present on the ACS images.

3.3  Criteria for identification as a merger

The above selection criteria result in five "samples" listed in Table 1, in which each of the three SDSS-based samples is associated with one of the two COSMOS samples. In order to preserve balance in the experiment and to make the visual classification task manageable, we decided to limit the number of COSMOS galaxies so as to match in number each of the available samples of SDSS galaxies. These were picked randomly from a larger sample of 6397 COSMOS galaxies at $z = 0.7$ and 1039 at $z = 1.2$. In the final classification there are thus essentially equal numbers of real high redshift COSMOS galaxies and added SDSS galaxies, except for the $z = 1.2$ sample where we post facto deleted some objects from the SDSS sample to match the apparent $I_{AB} < 24.0$ COSMOS limit (see section 3.2). For each of the final set of 7128 galaxies, a small "postage stamp" 10x10 arcsec$^2$ centered on each galaxy was extracted. These were then shuffled into a random order and visually inspected one by one. Galaxies with clear evidence of tidal tails and/or highly distorted images suggestive of a recent or ongoing merger event were classified as "mergers". Other objects that were not convincing enough to be classified as mergers, but which nevertheless appeared somewhat disrupted in their morphology, were classified as "distorted".

It should be noted that our visual merger classification scheme does not identify objects whose morphologies are not distorted or which otherwise lack features indicative of an interaction. This may be the case for example in the merging of two elliptical galaxies (Barnes & Hut 1986), and the current study may well not be sensitive to these gas-poor mergers.

After the classification of all the real and simulated high redshift galaxies was completed, the *original* SDSS images were re-examined in order to identify (a) which, if any, of the simulated images had been classified as a merger due to the projection of a normal SDSS galaxy with an object already present on the COSMOS ACS images, and, conversely (b) how many SDSS galaxies, that are seen as mergers at low redshift, are not recognized as such on the simulated ACS images because of poorer S/N or other effects. As discussed below, both of these pieces of information are required to interpret the results of the visual classification in terms of a redshift evolution. We assume that all apparent mergers on the original SDSS images are in fact "real mergers". It should be noted that some of these SDSS mergers involve faint companions and thus it is to be expected that not all of them would be recognizable as mergers at high redshifts.

Table 1 summarizes the measurements made on these different samples, which are discussed below. It should be noted that those entries in the Table that are preceded by an "=" sign are derived from other observed quantities in the Table (see below).

4. RESULTS AND DISCUSSION

4.1  The evolution in the corrected merger rate to $z \sim 1.2$

Looking first at the $z \sim 0.7$ comparison, we note the following (see also entries in Table 1):

(a) There is no significant difference between the fraction of "unbrightened" and "brightened" samples that are visually classified as mergers (1.8% and 2.3%) and so we henceforth average these to 2.0%.
(b) The bulk of SDSS objects that are classified as mergers on the simulated high redshift images are not in fact real mergers, but have been (mis-)classified as such because of chance projections with unrelated objects that were already present on the ACS images. This is despite the requirement for clear evidence of tidal distortions etc. in order to classify an object as a merger. In both the "brightened" and "unbrightened" $z \sim 0.7$ SDSS simulations, we find that between 1.5-2.0 % of all added objects are wrongly classified as mergers due to superpositions with unrelated objects. These "fake mergers" therefore comprise 80-90% of all the apparent mergers in the simulated SDSS images.
(c) Since 1.5-2.0% of randomly added "normal" SDSS objects at $z = 0.7$ appear as mergers because of chance superpositions, then it is reasonable to conclude that the same fraction of "normal" COSMOS objects at $z = 0.6-0.8$ will probably have been similarly misclassified.
(d) The observed fraction of real COSMOS objects that are classified as mergers is 4.1%. Of this 4.1%, 40% (i.e. 1.5-2.0% of the full sample) are presumably spurious by the logic of (c). This is lower than the 80-90% for SDSS because the total number of merger candidates is higher. Therefore the fixed number of spurious mergers

        comprises a smaller percentage of all the merger candidates. As a result, more than a half (~60%) of the COSMOS merger candidates are likely to be "real" mergers.

(e)      Only 6 (19%) of the real SDSS mergers, i.e. of the 32 systems that were identified by us in the original un-degraded SDSS images, are in fact recognized as mergers in the simulated COSMOS ACS images. This is because of S/N degradation, surface brightness dimming and so on. Another 12 were classified as "distorted", but the remainder (44%) had disappeared altogether in the sense of being visually classified as normal galaxies.

Our analysis highlights the difficulties in the visual classification of high redshift galaxies, even in relatively bright galaxies ($I_{AB} < 24$) and modest redshifts ($z \sim 0.7$), both in not recognizing real mergers and in wrongly identifying chance superpositions as mergers, and emphasizes the need to do careful simulations of the type presented here before interpreting the visual classification of images of galaxies at high $z$.

    Our simulations allow us to correct for both of these observational effects: the fraction of COSMOS galaxies at high redshift undergoing recognizable mergers reduces from 4.1% to 2.4%, to be compared with the 0.32% of local SDSS galaxies whose merger activity is still recognizable at high redshifts. We thus conclude that there is an increase in the fraction of galaxies undergoing mergers, between $z \sim 0$ and $z \sim 0.7$, of approximately a factor of eight (0.32% to 2.4%). If we parameterize this increase as $(1+z)^m$, then we find $m \sim 3.8 \pm 1.2$. This is shown in Fig. 2. The formal uncertainty in this estimate is dominated by the relatively small number of real SDSS mergers that remained detectable at high redshift (i.e. six of the original input sample of 1813).

    Our results indicate a relatively rapid increase in the fraction of mergers between redshift $z \sim 0.0$ and $z \sim 0.7$. However, as noted above, at $z \sim 0.7$ we may well be identifying only 20% of all mergers due to the effects of image degradation and noise etc.. This further compounds in our analysis the usual uncertainties that are present regarding the timescales over which a given merger event would be morphologically classified as a merger.

    At the higher redshifts, the uncertainties become even larger. In fact, none of the actual SDSS mergers is recognized at $z = 1.2$ as a merger in the simulated images and all of the apparent mergers in the SDSS sample are in fact superpositions (or, in two cases, are mis-classified as a merger due to S/N degradation). Perhaps surprisingly, the fraction of SDSS objects that are misclassified as mergers due to superpositions is actually smaller (0.65%) at $z \sim 1.2$ than at $z \sim 0.7$, even though the number density of foreground or background objects is by construction the same (since both were added to actual ACS images). Our impression is that this is because the SDSS u-band images appear less radially extended than the g-band images, presumably because of surface brightness effects, which reduces the probability of a chance overlap which may be mistaken for a tidal feature. In contrast, the fraction of apparent mergers that are seen in the real COSMOS galaxies continues to increase, up to 5.2%, and we suspect that the great majority of these are in fact real mergers.

    We can correct the observed fraction at $z = 1.2$ for superposition, as above, down to 4.6%, but without a zero-redshift anchor from the local sample (since no real SDSS mergers were recognized at $z = 1.2$) we cannot formally compute a value for $m$. However, we can still compare this 4.6% fraction with our estimate at $z = 0.7$ if we regard it as a lower limit on the basis that the detectability of known mergers is unlikely to be higher at $z = 1.2$ than at $z = 0.7$

(see Fig.2). We take this lower limit as providing further support for an increase in the real merger fraction with increasing redshift from $z \sim 0.7$ to $z \sim 1.2$.

4.2     The absolute magnitude distribution of mergers

The above analysis considers the various samples without regard to the absolute magnitudes of the individual galaxies beyond requiring that the limits of the absolute magnitude distributions be similar. Variations in the identified merger fraction with absolute magnitude are of potential interest both astrophysically and as an indicator of possible systematic uncertainties or failures in the adopted methodology.

Fig. 3 shows the distribution of absolute magnitudes for the various samples at $z = 0.7$ in cumulative form. It is clear that the apparent mergers from COSMOS sample do not trace the distribution of luminosities of all COSMOS objects, and are biased towards brighter objects. This likely represents a straightforward bias in the ability to recognize mergers as we approach the limiting magnitude of $I_{AB} \sim 24$ – recall that the photometry is based on the ACS images and should not artificially boost the brightness of close pairs of galaxies. Looking at the SDSS samples in Fig.3, it is noticeable that the same effect is visible in the comparison of the luminosity distribution of the "real" SDSS mergers with those that are successfully recovered at high redshift. The effect is weaker in the luminosity distribution of the apparent mergers in the simulated SDSS sample, consistent with the idea that these are often not real mergers but chance superpositions. Again, note that the luminosities of SDSS galaxies are computed before addition to the ACS images, so that the effect of superposed objects is by construction not present. Given that a much higher fraction of the COSMOS mergers are "real" it is not surprising that the comparison of the COSMOS distributions looks more like that of the real SDSS mergers than the apparent SDSS mergers.

It is also not surprising that there is a trend in the detectability of a merger with luminosity. We believe that the distribution of luminosities between the parent SDSS and COSMOS samples is sufficiently similar (the heavy dashed lines in the two panels of Fig. 3) that this will not introduce a spurious trend with redshift.

However, this serves to emphasize that the fraction of galaxies that are "identifiable" mergers, as derived in this paper, will be a lower limit to the fraction of galaxies that are actually merging, which may be substantially higher. The strength of our approach relies on recognizing that our selection will be highly incomplete, but ensuring as far as we can that this incompleteness is not redshift dependent.

4.3     The colors of mergers at $z \sim 0.7$

We compare the color distribution of COSMOS objects at $z \sim 0.7$ classified as mergers with those of the COSMOS sample as a whole, remembering that about 40% of the former are likely to be chance superpositions of objects. We use the observed $(B-I)_{AB}$ colors computed in 3 arcsec diameter apertures by Capak et al. (2006) from ground-based photometry obtained with the Subaru telescope (Taniguchi et al. 2006). We find that the color distribution of all merger candidates (Fig.4) is significantly skewed towards bluer objects. The Kolmogorov–Smirnov (K-S) statistical test confirms this with confidence level greater than 99%.

We can try to remove, statistically, the misclassified mergers that are in fact chance superpositions of unrelated objects, if we assume that these misclassified objects will have the same color distribution as other COSMOS galaxies. This is likely to be grossly true but to be incorrect in detail, depending on the colors and relative magnitudes of the contaminating objects and their location relative to the 3 arcsec aperture used to compute the colors. Testing this assumption would require additional simulations that are beyond the scope of the present paper. However, applying this assumption to subtract a scaled version of the overall color distribution, we obtain the modified histogram shown in Fig.4. This further accentuates the predominance of blue over red mergers.

It is natural to ascribe this result to enhanced levels of star-formation associated with the merger event, and one could consider this result to be evidence *against* dry mergers involving galaxies without gas. However, as noted above, our visual classification could be biased against mergers involving galaxies without a pre-existing cold disk, since these are currently thought to produce the strongest tidal features. New high-resolution N-body simulations involving realistic progenitors and orbital distributions seem however to indicate that strong tidal features, lasting several billion years, may also result from gas-free mergers (Carollo & Mayer, in preparation). We nonetheless caution against making this conclusion, at least for the time being.

4.4     Morphologies

Various non-parameteric descriptors of galaxy morphology have been developed over the last few years with the motivation of automatically classifying large numbers of faint galaxy images (see e.g. Abraham et al. 1996, Conselice et al. 2003, Lotz et al. 2006). Scarlata et al. (2006) have taken this approach one step further by developing ZEST, a galaxy classification scheme based on a principle component analysis based on multiple non-parametric morphological descriptors including elongation, which was applied to the same set of COSMOS galaxies analysed here.

Fig.6 shows the ZEST morphological classification of our galaxies as derived in Scarlata et al. (2006). It should be noted that, as described in that paper, these parameters are derived from images that have been "cleaned" of nearby companions since the goal was to classify the underlying galaxies. Comparing the ZEST classification of the merger candidates we find that 30% of ZEST irregulars are classified by us as merger candidates, a fraction that is much higher than that for other morphological types, which increases from 2%-5% from elliptical to bulge-less spiral (ZEST classes 1.0 to 2.3). Nevertheless the majority (63%) of our merger candidates have the underlying morphologies of regular elliptical and spiral galaxies as determined by ZEST. The interpretation of this statistic is made difficult by the fact that 40% of the COSMOS merger candidates are expected to be chance superpositions, for which we cannot, obviously, correct as we did before for the colors.

Turning to the non-parametric descriptors themselves, we have examined the distribution of the Concentration, Asymmetry, $M_{20}$ and Gini parameters measured by Scarlata et al. (2006) for the 115 merger candidates with those for the 2831 galaxies in the overall sample. The largest differences between these distributions are seen in the Asymmetry parameter, *A*. This is shown in Fig. 5 where we plot histograms of the two (normalized) distributions, the cumulative distributions and the fraction of mergers as a function of *A*. As might be expected, even though neighboring objects were cleaned before computation of *A*, the galaxies identified by us as mergers are generally more asymmetric in their underlying structure than the others, with typical values of A that are about 70% larger, over a wide range of percentiles of the distributions. Likewise, whereas very few of the galaxies with A < 0.2 are classified by mergers, we find that roughly a third to a half of those with $A > 0.2$ are so classified. A straightforward selection of objects in Scarlata et al. (2006) with $A \geq 0.2$ would include 60% of our merger classifications but would introduce an equal number of galaxies that we have not classified as mergers.

We do not see significant differences in the distribution of Concentration or Gini parameters, and only a weak effect in $M_{20}$. Lotz et al. (2006) have used the $M_{20}$-Gini diagram as a diagnostic to identify mergers in high redshift galaxy samples on the basis of the location of low redshift ultra luminous infrared galaxies (ULIRGs) on this diagnostic diagram (Lotz et al. 2004). In our own sample, we find that only about 6% of our visually classified mergers lie in the region of the $M_{20}$-Gini plane used by Lotz et al. (see Fig.7) and that the majority overlap with the locus of normal galaxies, consistent with the ZEST classification shown in Fig.6.

4.5     Evolution in the fraction of distorted galaxies

Our analysis above has focused on objects classified as "mergers", even though, as we have argued, many apparent mergers are likely not mergers at all but arise through chance superpositions. During the visual inspection of the images, we also noted those objects with clearly distorted morphologies that however did not merit the merger classification. Although our "distorted" classification is even more subjective than the "merger" classification, the fact that the classification was done "blind" and was based on images that take proper account of the redshift-dependent observational effects and biases described above, makes it still worthwhile to compare the fraction of objects in each sample that are classified as distorted. We do not correct for "superpositions" but can compare the fractions at different redshifts.

The fraction of these "distorted" objects is shown on Fig. 2. The increase with redshift is less steep than for the increase in the corrected merger fraction with $m = 1.8 \pm 0.2$ for $0 < z < 0.7$ and $m = 1.8 \pm 0.3$ for $0 < z < 1.2$. Interestingly, this is similar to the increase in apparent merger fraction before correction for the effects of superposition (see Fig. 2). The "distorted" objects no doubt represent a range of phenomena. As noted above, some of them (of order 1/3 in the simulated SDSS sample) are likely to be mergers that would be recognized as such with better imaging data. The remainder may represent more gradual "accretion" processes.

4.6     Discussion

Our morphological analysis described above has highlighted the difficulties of studying the morphological evolution of galaxies as a function of redshift and has emphasized the need to carry out careful simulations of the type described here. Not least, the effects of chance superpositions have been greater than we expected, and have demonstrated the need to include realistic "background" fields when simulating high redshift galaxies, rather than the common practice of using blank regions of sky.

The change in the merger fraction with redshift that we have derived from a purely morphological classification, $m \sim 3.8 \pm 1.2$, is in the upper half of the range published in the literature. It is quite consistent with the early CFRS-based morphological estimate of Le Fèvre et al. (2000) although the corrected merger fractions are much lower (see also Bundy et al.

2004). Probably the most obvious empirical discrepancy is with Lotz et al. (2006) who adopted an automated classification scheme discussed in the previous section and who derived a low value of $m = 1.1 \pm 0.6$.

We note that our own estimate of $m$ is also consistent with the independent analysis of galaxy pair counts in the COSMOS field by Karteltepe et al. (2006). It is to be hoped that future analyses of the very large COSMOS dataset in terms of infrared-selected starbursts, spectroscopic kinematics and the galaxy mass function will further illuminate the role that mergers play in the assembly of galaxies over cosmic time.

5. SUMMARY

Simulations of nearby ($0.015 < z < 0.025$) SDSS galaxies, that reproduce, as accurately as possible, the appearance that they would have if observed at $z \sim 0.7$ and $z \sim 1.2$ in COSMOS ACS images, have been used to examine the redshift-dependent observational effects that will affect the visual classification of galaxies as mergers at high redshift. These simulations include the well-known surface brightness dimming, the effects of band-pass shifting and changing spatial resolution, as well as the proper ACS-COSMOS noise properties. Since the SDSS galaxies are added to real COSMOS ACS images without regard to the location of pre-existing galaxies, the simulations also accurately mimic the effects of chance superpositions of high redshift galaxies with unrelated foreground or background objects. Both "brightened" and "unbrightened" versions of the SDSS galaxies are made, the former increasing the surface brightness at high redshift by an amount $\Delta\mu = z$.

These simulated images of local galaxies as seen at $z \sim 0.7$ and $z \sim 1.2$ have been compared with a roughly equal number of images of real galaxies at these two redshifts, selected from the COSMOS sample on the basis of their photometric redshifts. In particular, a completely "blind" identification of galaxies that appear to be mergers, and also those which are otherwise distorted but not clearly a merger candidate, has been undertaken. The absolute magnitude range of the SDSS galaxies is chosen to match that of the real COSMOS galaxies, regardless of whether surface brightness evolution is applied or not.

It is found that purely observational effects play a significant role in modifying the appearance of galaxies at high redshift and in the conclusions that may be drawn about mergers in the distant Universe. These make known mergers in the local Universe harder to recognize and produce apparent, but spurious, mergers through the superposition of unrelated galaxies. Whereas we classified about 1.8% of SDSS galaxies to be mergers on the original SDSS images, less than one in five of these (i.e. 0.3% of all galaxies) are still recognized as mergers when they are transformed to $z = 0.7$ and added to the COSMOS ACS images. Conversely, we find that 1.7% of all normal SDSS galaxies that added to the ACS images are classified as "mergers" at high redshift simply due to the superposition of unrelated foreground or background objects.

The fraction of real COSMOS galaxies that were visually classified as mergers is however higher than for the transformed SDSS objects, at about 4% at $z \sim 0.7$, and 5.2% at $z \sim 1.2$. If we correct these fractions for the superposition fraction (which we assume is the same as in the SDSS simulations) then we infer that most of the COSMOS mergers are nevertheless real (i.e. 60% at $z = 0.7$ and 90% at $z = 1.2$).

Comparing this corrected number of real mergers seen in COSMOS at $z = 0.7$ with the number of mergers in SDSS that are detectable at high redshift, we find that the fraction of galaxies undergoing mergers increases as $(1+z)^{3.8\pm1.2}$ to $z \sim 0.7$, with the uncertainty dominated by the poor statistics of mergers in the current SDSS DR4 sample. Because we actually detect no real SDSS mergers at $z = 1.2$, we can not compute $m$ to this redshift. However, if we regard the corrected number of real mergers seen in COSMOS at $z = 1.2$ as a lower limit when compared with the detectability threshold at $z = 0.7$ (rather than correcting to a new detectability threshold), then this lower limit clearly continues the steep rise in redshift out to $z = 1.2$.

We find that, compared with the parent COSMOS sample at $z \sim 0.7$, the objects classified as mergers are significantly bluer, especially when a statistical correction is made for the spurious mergers caused by projection. In fact, almost no mergers with red colors are found. It would be premature to conclude that this rules out dry merging between galaxies without a gas disk because such mergers might not have produced the strong tidal features that we use to identify mergers.

Comparing the distributions of non-parametric morphological parameters of the merger candidates with the parent population, we find that the mergers are significantly more asymmetric, but do not have different distributions in Concentration, Gini, with only a small effect in $M_{20}$. Galaxies that are classified as irregulars by the automated ZEST classifier (Scarlata et al. 2006) are much more likely to have been classified by mergers by us, but most merger candidates (63%) have underlying morphologies that are nevertheless classified as those of regular elliptical or spiral galaxies. Perhaps as result, less than 10% of our merger candidates lie in the region of the M20-Gini plane that has been used by Lotz et al. (2004, 2006) to isolate mergers.

The fraction of less extreme distorted morphologies, which might be associated with "minor" mergers involving smaller companion galaxies, is higher and also shows a rise with redshift, although the increase in $(1+z)^m$ is shallower with $m \sim 1.8$.

Acknowledgements

The HST COSMOS Treasury program was supported through NASA grant HST-GO-09822. We gratefully acknowledge the contribution of the entire COSMOS collaboration consisting of more than 70 scientists. It is a pleasure to acknowledge the excellent services provided by the NASA IPAC/IRSA staff in providing online archive and server capabilities for the COSMOS datasets. More information on the COSMOS survey is available at http://www.astro.caltech.edu/~cosmos. P. Kampczyk and C. Scarlata acknowledge support from the Swiss National Science Foundation.

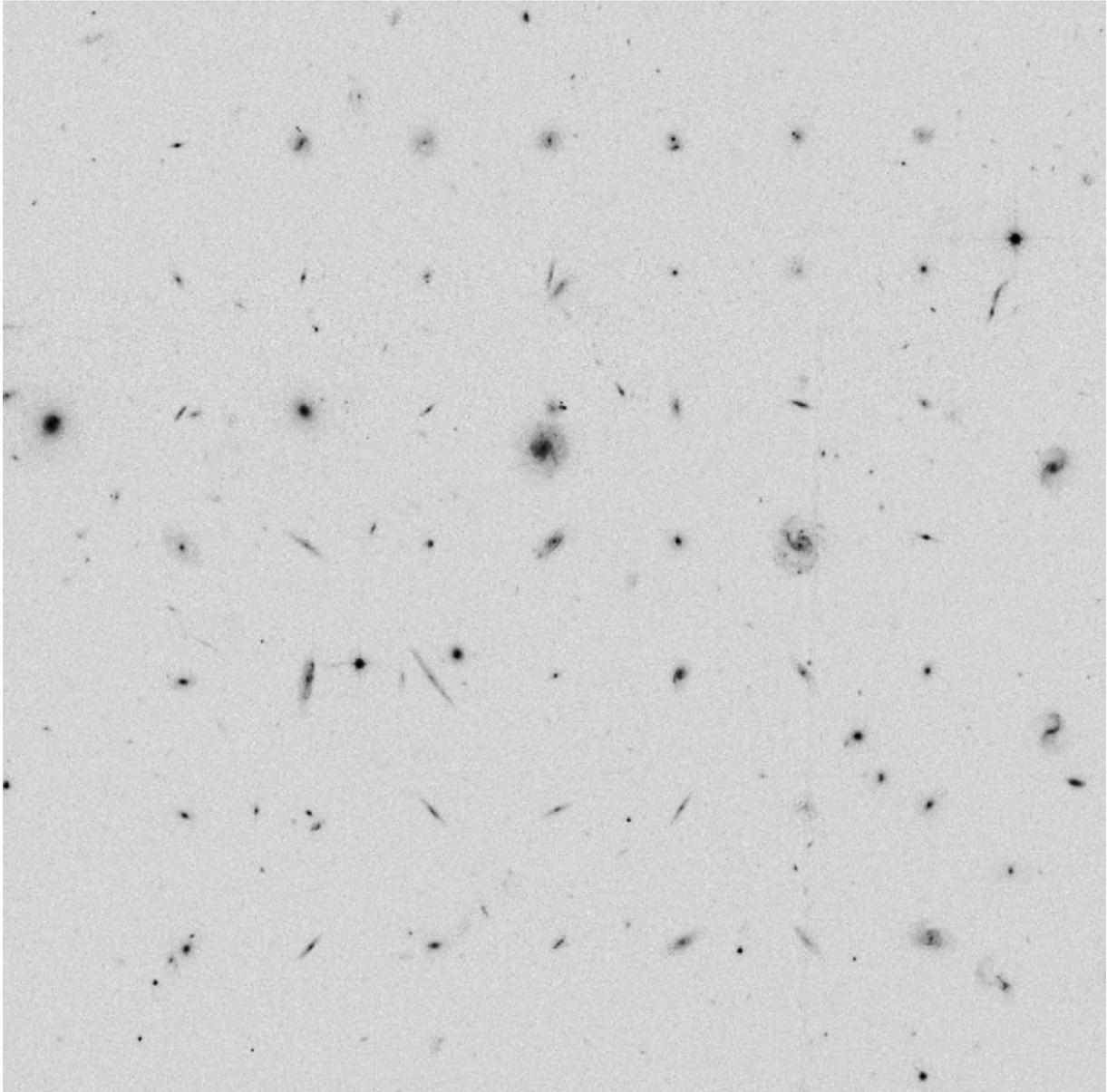

Fig. 1: Central portion of one COSMOS ACS tile containing 49 SDSS galaxies processed to appear as they would at z = 0.7 and then pasted into the image in a 7 x 7 positions grid, each position separated by 10 arcsec. No attempt is made to place the simulated galaxies on regions of blank sky, thereby simulating the same statistical superposition of foreground and background galaxies as real galaxies at this redshift.

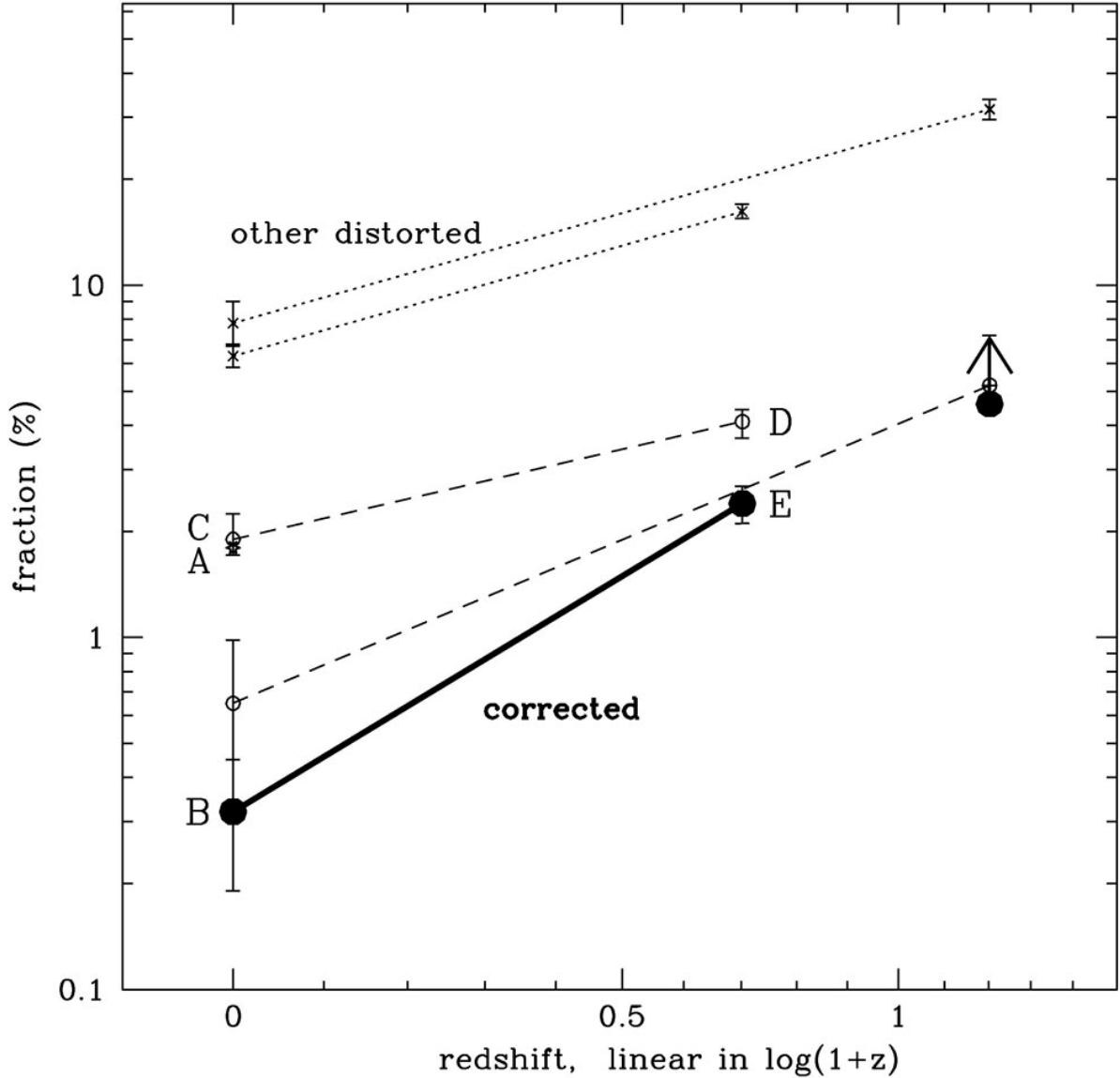

Fig. 2. Summary of observations: Each high redshift point, representing a measurement of real COSMOS galaxies at this redshift, is associated with a low redshift point, corresponding to the appropriate SDSS simulation, to which it is joined by a line. For the $z = 0.7$ measurements and simulations, the relevant points are labeled as follows: (A) the fraction of SDSS galaxies classified as mergers on the original SDSS images; (B) the fraction of these that are still observable at $z = 0.7$ despite observational degradations; (C) the fraction of SDSS objects that are in fact classified as mergers in the $z = 0.7$ simulations, the increase from B being due to the effects of unrelated superposed objects on the ACS images; (D) the fraction of real COSMOS galaxies classified as mergers at $z = 0.7$; (E) the fraction of actual mergers in COSMOS once the presumed superpositions are removed. The best estimate for the change in merger fraction is the line BE, i.e. $(1+z)^{3.8\pm1.2}$. An omission of one or both of the observational effects considered in this paper, i.e. image degradation and superposition of unrelated galaxies, would result in the apparent evolution vectors AD, BD, or AE. The light dotted lines are the same but for the additional objects classified as "distorted"; these have not been corrected for the effects discussed above. Because no real SDSS mergers were still be detectable at $z = 1.2$, we cannot produce a corrected low redshift point for mergers at this redshift, and have therefore plotted the corrected high redshift point as a lower limit on the grounds that the observability of mergers at $z = 1.2$ can be no better than at $z = 0.7$.

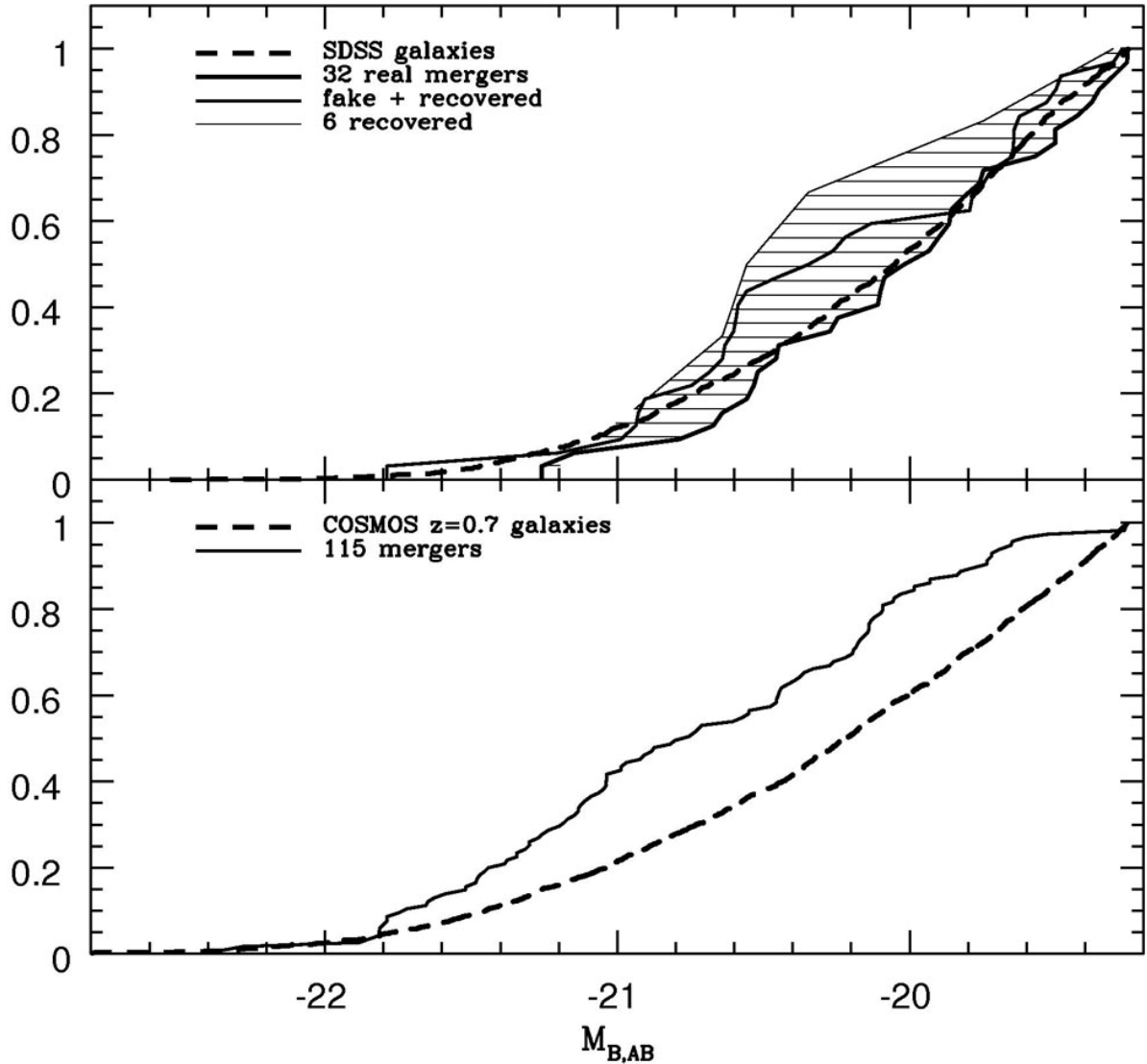

Fig. 3. Cumulative distributions of absolute magnitude in the simulated $z = 0.7$ SDSS "brightened" sample (top) and the $z = 0.7$ COSMOS sample (below). The overall luminosity distributions of the two samples (dashed lines) are similar. The hatched area in the upper panel highlights the fact that the brighter "real" SDSS mergers are easier to detect at high redshift in our simulations. The bias in the apparent mergers is smaller, probably reflecting the fact that most of these are in fact chance superpositions. The apparent COSMOS mergers are also brighter than the parent sample, presumably for the same reason.

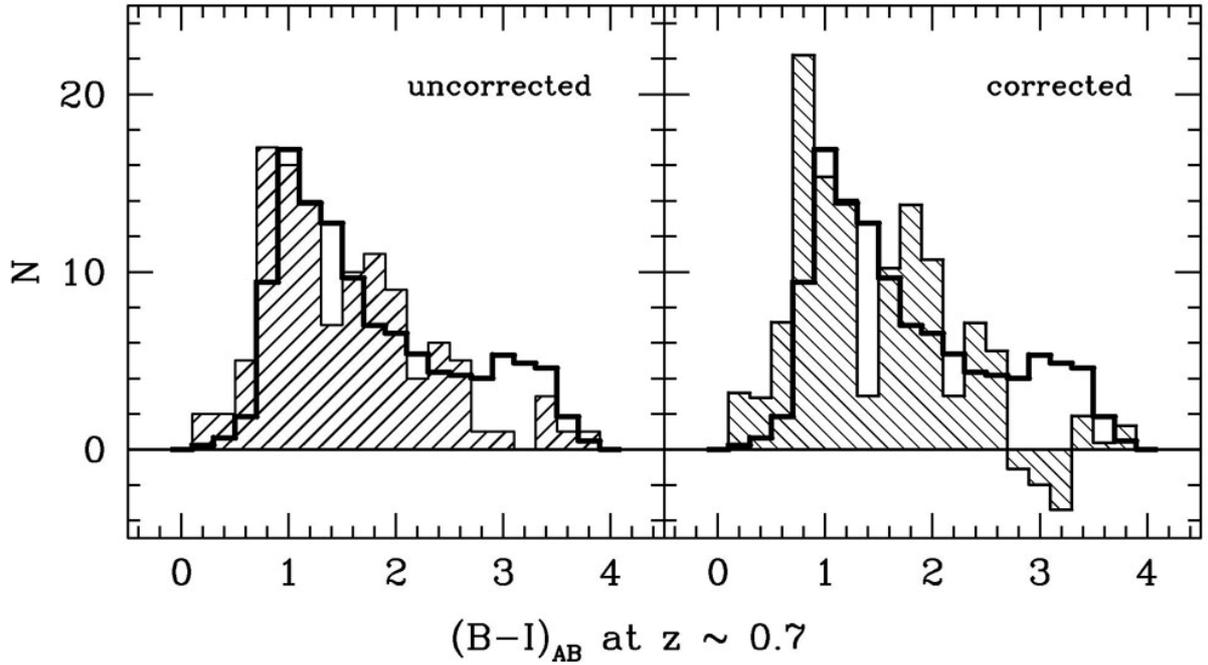

Fig. 4. (Left) The observed color distributions of the apparent 115 mergers in the $z \approx 0.7$ COSMOS sample (shaded histogram – left panel) compared to the normalized distribution of colors of all 2831 COSMOS galaxies at this redshift in the $I_{AB} \leq 24$ catalog of Scarlata et al. (thick line – left panel). These distributions are different at the 99% confidence level and indicate that mergers are on average bluer than the other galaxies. (Right panel) Taking into account the effect of contamination caused by random projections results in even greater (99.9% confidence level) difference between distributions of all COSMOS galaxies (thick line) and mergers corrected for contamination (shaded histogram).

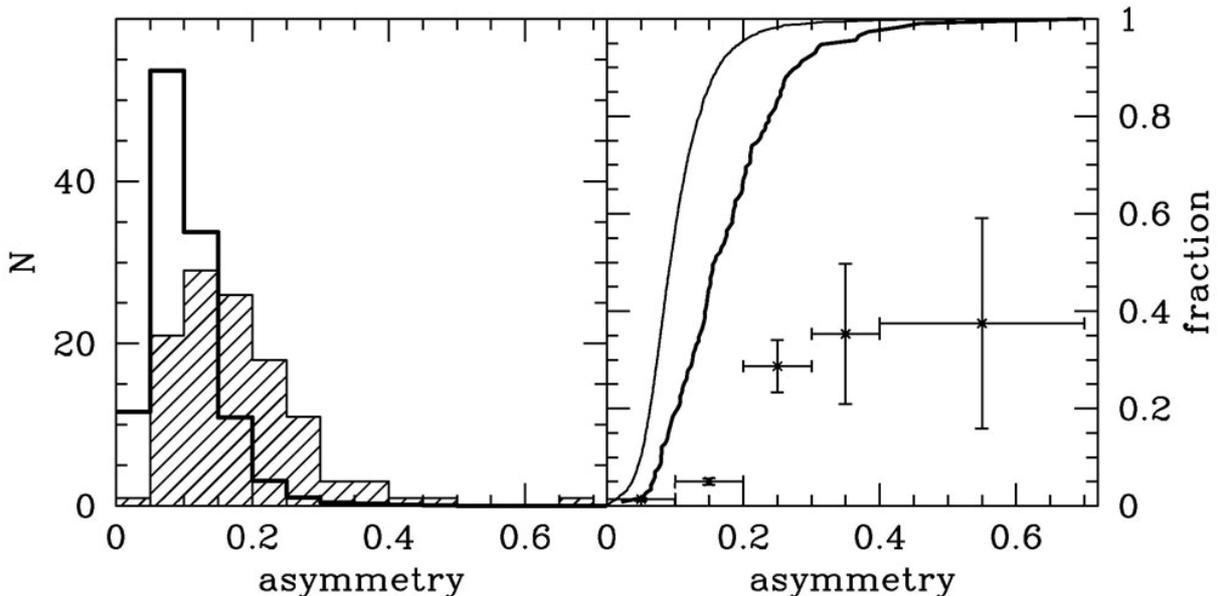

Fig. 5. (Left) Distribution of asymmetry parameter for the apparent mergers in the $z \approx 0.7$ COSMOS sample (shaded histogram) compared to the normalized distribution of 2831 COSMOS galaxies at this redshift (thick line). Mergers are significantly more asymmetric than galaxies in general. (Right panel) Cumulative distributions of Asymmetry parameter for all galaxies (thin line) and mergers (thick line). On average mergers are 70%

more asymmetric than all galaxies. Points with error bars represent the fraction of mergers in given Asymmetry bin. The fraction of mergers increases significantly with higher asymmetry.

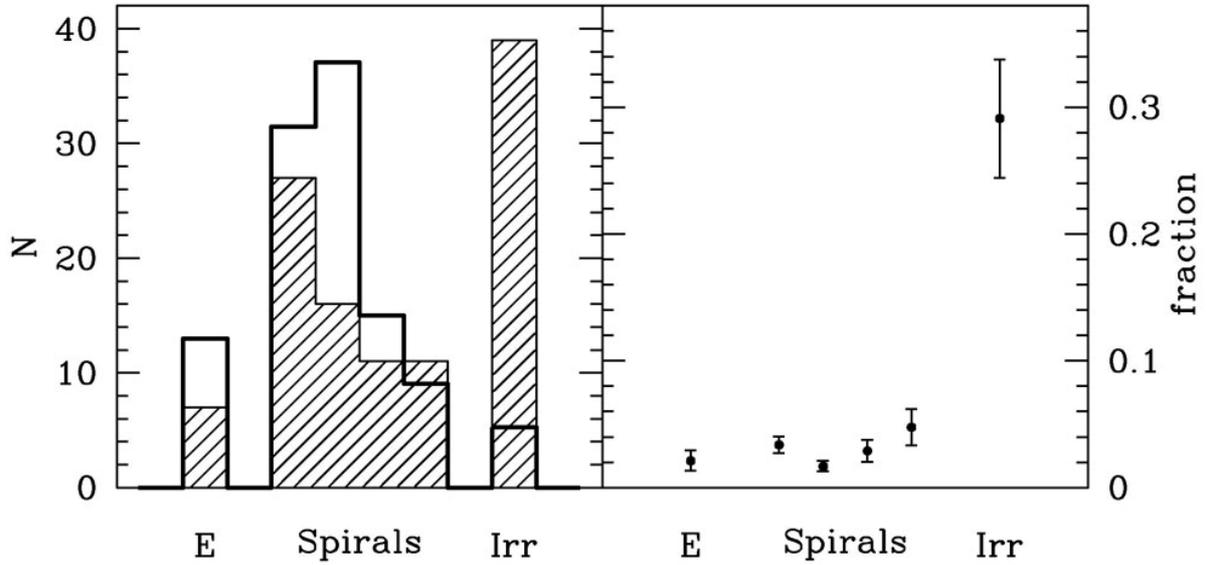

Fig. 6. (left) Normalized distribution of ZEST morphological types from Scarlata et al. 2006 for the sample of all 2831 COSMOS galaxies at $z \sim 0.7$ (heavy line) compared with those for the 115 galaxies classified as mergers in the present analysis. The spiral galaxies were divided into 4 bins from bulgeless to bulge dominated disks (left to right) corresponding to ZEST classes 2.0 to 2.3. (right) The fractions of visually classified mergers in a given ZEST morphological class.

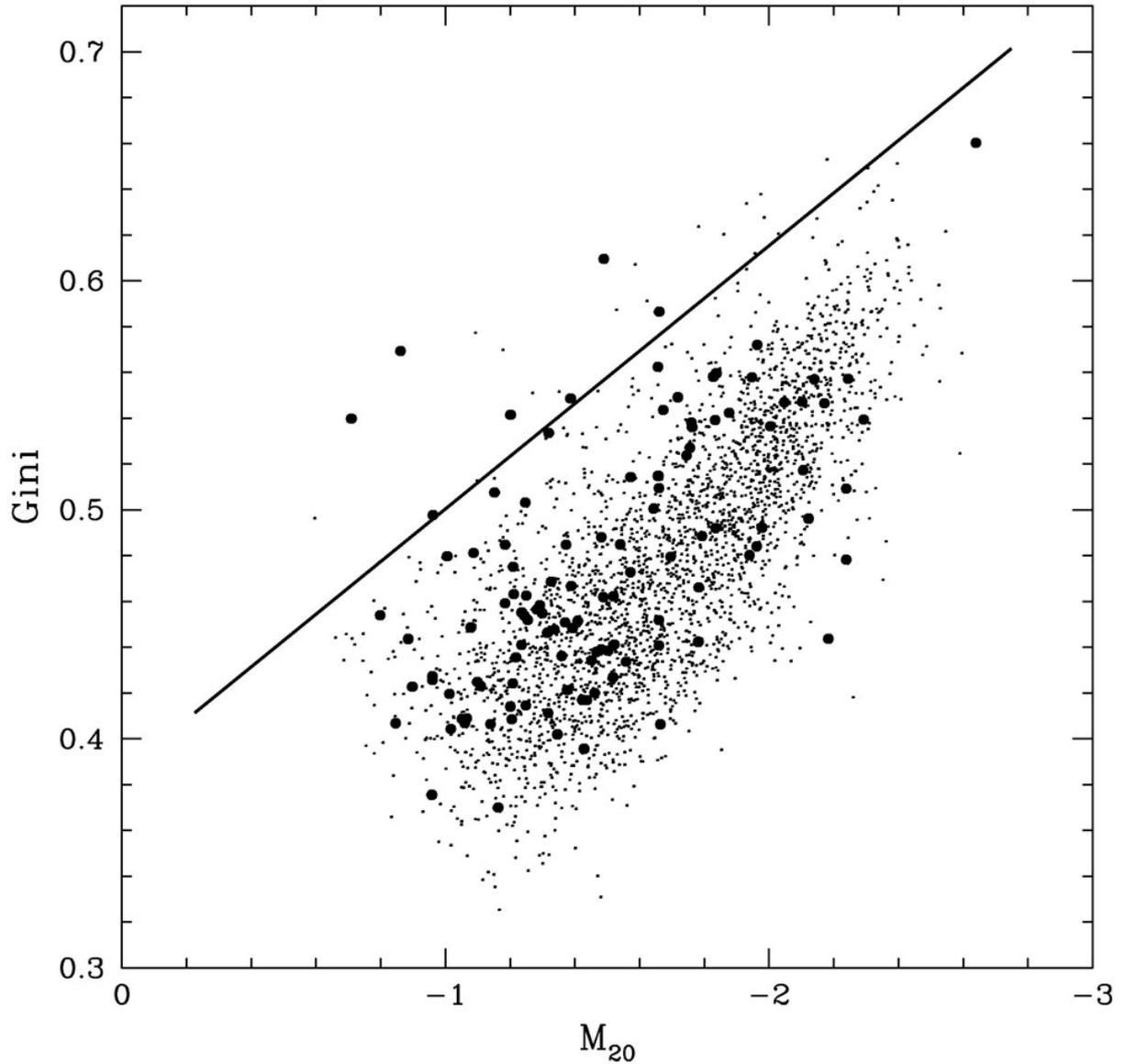

Fig. 7. Gini v. $M_{20}$ parameters for the sample of 2831 COSMOS galaxies from redshift $0.6 \leq z \leq 0.8$ and the 115 galaxies visually identified as mergers, represented by the larger points. The plotted line shows the selection criterion for mergers used in Lotz et al. (2006): $G > -0.115 M_{20} + 0.384$. In our sample less than 10% of mergers lie in this area.

**Table 1: Measurements**

| | Sample | $M_{B,AB}$ | z range of input galaxies | N | Number of apparent mergers on real or simulated ACS images | Number of apparent mergers that are due to chance projections | Number of real mergers after removal of contaminants | Number of mergers on original SDSS images | Number of "distorted" objects |
|---|---|---|---|---|---|---|---|---|---|
| $z \sim 0.7$ | SDSS *g*-band "unbrightened" | $\leq$ -19.25 | 0.015 – 0.025 | 1018 | 23 (2.3%) | 20 (2.0%) | 3 (0.29%) | 16 (1.6%) | 56 (5.5%) |
| | SDSS *g*-band "brightened" | $\leq$ -18.55 | 0.015 – 0.025 | 1813 | 33 (1.8%) | 27 (1.5%) | 6 (0.33%) | 32 (1.8%) | 123 (6.8%) |
| | COSMOS | $\leq$ -19.25 | 0.7 – 0.8 | 2831 | 115 (4.1%) | = 47 (1.7%) | = 68 (2.4%) | — | 460 (16.2%) |
| $z \sim 1.2$ | SDSS *u*-band "brightened" | $\leq$ -19.55 $m_I \leq 24$ | 0.015 – 0.025 | 611 | 4 (0.65%) | 4 (0.65%) | 0 | — | 48 (7.8%) |
| | COSMOS | $\leq$ -20.80 | 1.18 – 1.24 | 733 | 38 (5.2%) | = 4 (0.55%) | = 34 (4.6%) | — | 232 (31.6%) |